\input{psfig}

\documentstyle[floats,aps,aipbook]{revtex}
\begin{document}
\title{Results on Hard Diffractive Production}
\author{Konstantin Goulianos\footnote{Presented at the 10$^{th}$
Topical Workshop on Proton-Antiproton Collider Physics, 9-13 May 1995,
Fermi National Accelerator Laboratory.}}
\address{The Rockefeller University\\New York, NY 10021, USA\\
(The CDF Collaboration)}
\maketitle
\begin{abstract}
The results of experiments at hadron colliders probing the structure of the
pomeron through hard diffraction are reviewed. Some results on
deep inelastic diffractive scattering obtained at HERA are
also discussed and placed in perspective.
By using a properly normalized pomeron flux factor in single diffraction
dissociation, as dictated by unitarity, the pomeron emerges as a
combination of valence quark and gluon color singlets in a ratio
suggested by asymptopia.
\end{abstract}

\section{Introduction}
In this paper we review results obtained in studies of hard diffractive
production at hadron colliders and deep inelastic diffractive
scattering at HERA
and draw conclusions about the QCD structure of the pomeron.
The results on diffractive $W$ and dijet production are
presented on behalf of the CDF Collaboration.

The phenomenology associated with extracting information
on the pomeron structure
from these studies relies on Regge theory and factorization.
\begin{figure}[h,t,b]
\vspace*{-0.5in}
\centerline{\psfig{figure=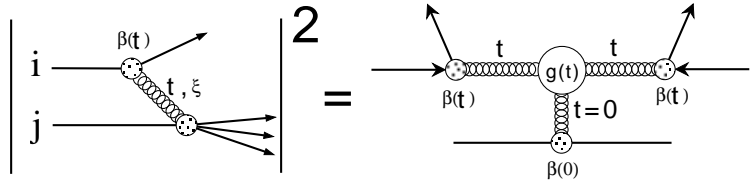,width=6in}}
\vspace*{-6.35in}
\caption{The triple-pomeron amplitude for single diffraction dissociation.}
\label{triple-pomeron}
\end{figure}

The cross section for
single diffraction dissociation
in Regge theory has the form (see Fig.~\ref{triple-pomeron})
\begin{equation}
\frac{d^2\sigma_{sd}^{ij}}{dtd\xi}=
\frac{1}{16\pi}\;\frac{\beta_{i{\cal{P}}}^2(t)}
{\xi^{2\alpha(t)-1}}\left[ \beta_{j{\cal{P}}}(0)\,g(t)
\;\left(\frac{s'}{s'_0}\right)^{\alpha(0)-1}\right]=f_{{\cal{P}}/i}(\xi,t)\;
\sigma_T^{{\cal{P}}j}(s',t)
\label{diffractive}
\end{equation}
where ${\cal{P}}$ stands for pomeron, $s'$ is the s-value in the
${\cal{P}}-j$ reference frame, $s'_0$ is a constant conventionally set to
1 GeV$^2$,
$\xi=s'/s$ is the
{\mbox{Feynman-$x$}} of the pomeron in hadron-$i$, and $\alpha(t)$ the pomeron
trajectory given by
$\alpha(t)=\alpha(0)+\alpha't=1+\epsilon+\alpha't$.
The term in the square brackets
is interpreted as the total cross section of the pomeron on hadron-$j$,
$\sigma_T^{{\cal{P}}j}(s',t)$, where $g(t)$ is the
``triple-pomeron coupling constant".
This interpretation leads naturally to viewing single diffraction
as being due to a flux of pomerons, $f_{{\cal{P}}/i}(\xi,t)$, {\em emitted}
by hadron-$i$ and interacting with hadron-$j$.

Assuming that factorization holds in {\em hard} processes, as it does in soft
\cite{KG1},
Ingelman and Schlein (IS) used the pomeron flux from Eq.~\ref{diffractive}
to calculate high-$P_T$ jet production in $p\bar p$ single diffraction
dissociation \cite{IS}. Their calculation was followed by the
discovery of diffractive dijets by UA8 \cite{UA8}. However, the measured
dijet rate turned out to be substantially lower than the rate predicted by the
IS model \cite{UA8-EPS}. One possible explanation for this result is that
the {\em virtual} pomeron does not obey the momentum sum rule
\cite{UA8-EPS,DL2}.  A more physical explanation,
in which the pomeron {\em obeys}
the momentum sum rule, is offered by extending the IS model to
interpret the pomeron flux as a probability density for finding a pomeron
inside hadron-$i$ and
{\em renormalizing} it
so that it is not allowed to exceed unity \cite{KG2}.
As we shall see below,
using a renormalized pomeron flux lowers the predicted rates and brings
the UA8 results into agreement with
the momentum sum rule.

The pomeron flux renormalization
procedure was  proposed in order to
unitarize single diffraction dissociation.  Without renormalization,
the  $p\bar p$ single diffractive
cross section rises much faster than that observed,
reaching the total cross section and therefore violating unitarity
at the TeV energy scale.
The renormalized flux is given by
\begin{equation}
f_N(\xi,t)=\frac{f_{{\cal{P}}/i}(\xi,t)d\xi dt}{N(\xi_{min})}
\label{FN}
\end{equation}
$$N(\xi_{min})\equiv \left\{ \begin{array}{ll}
1&\mbox{if $N(\xi_{min})\leq 1$}\\
{\int_{\xi_{min}}^{0.1}d\xi \int_{t=0}^{\infty} f_{{\cal{P}}/i}(\xi,t)\;dt}
&\mbox{if $N(\xi_{min})> 1$}
\end{array}\right. $$
where $\xi_{min}$=$(1.5\; \mbox{GeV}^2/s)$ for $p\bar p$ soft single
diffraction \cite{KG2}.
Below, experimental results on hard diffraction
will be compared with predictions obtained both with the standard and with the
renormalized pomeron flux.
\section{Hard diffraction at hadron colliders}
\begin{figure}[h,t,b]
\centerline{\psfig{figure=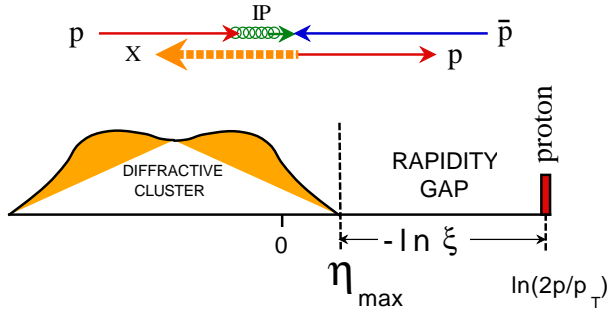,width=6in}}
\vspace*{-5.85in}
\caption{Pseudorapidity distribution of particles in diffraction dissociation.}
\label{rapgap}
\end{figure}

Events are tagged
as diffractive either by the detection of a high-$x_F$
(anti)proton, which presumably ``emitted" a small-$\xi$ pomeron,
or by the presense of a rapidity gap at one end of the kinematic region,
as shown in Fig.~\ref{rapgap}.
\subsection{The UA8 experiment}
UA8 pioneered hard diffraction studies
by observing high-$P_T$ jet production in the
process $p+\bar{p}\rightarrow p+Jet_1+Jet_2+X$ at the CERN $Sp\bar{p}S$
collider at $\sqrt{s}=630$ GeV.  Events with two jets of
$P_T>8$ GeV were detected
in coincidence with a high-$x_F$ proton,  whose momentum and angle
were measured in a forward ``roman pot" spectrometer. The event sample spanned
the kinematic range
$0.9<x_p<0.94$ and $0.9<|t|<2.3\;\mbox{GeV}^2$.
By comparing the $x_F$ distribution of the sum of the jet
momenta in the pomeron-proton rest frame with Monte Carlo
distributions generated with
different pomeron structure functions, UA8 concluded \cite{UA8}
that the partonic
structure of the pomeron is $\sim 57$\% {\em hard} [$6\beta (1-\beta)$],
$\sim 30$\% {\em superhard} [$\delta (\beta)$],
and $\sim 13$\% {\em soft} [$6(1-\beta)^5$].
However,  the measured dijet production rate was found to be smaller than that
predicted for a hard-quark(gluon) pomeron obeying the momentum sum rule by a
``discrepancy factor" of $0.46\pm0.08\pm0.24$  ($0.19\pm0.03\pm0.10$)
\cite{UA8-EPS}. Using the renormalized pomeron flux, the
discrepancy factor becomes
$1.79\pm 0.31\pm 0.93\;(0.74\pm 0.11\pm 0.39)$  \cite{KG2}, which is
consistent with the momentum sum rule.

\subsection{Diffractive W's in CDF}
The quark content of the pomeron
can be probed directly with diffractive $W$ production, which to leading order
occures through $q\bar{q}\rightarrow W$.  A hard gluonic pomeron can also
lead to diffractive $W's$ through $gq\rightarrow Wq (\rightarrow W+Jet)$,
but the rate for this
subprocess is down by a factor of order $\alpha_s$. The ratio of diffractive
to non-diffractive $W^{\pm}(\rightarrow l^{\pm}\nu )$ production has been
calculated by Bruni and Ingelman (BI) \cite{BI}
to be $\sim 17$\% ($\sim 1$)\% for a
hard-quark(gluon), and $\sim 0.4$\% for a soft-quark pomeron structure.
Thus, diffractive $W$ production is mainly sensitive to the hard-quark
component of the pomeron structure function. However,
using the renormalized pomeron flux
lowers the hard-quark prediction down to 2.8\%  \cite{KG2}.

\begin{figure}[h,t,b]
\vspace*{-0.6in}
\centerline{\psfig{figure=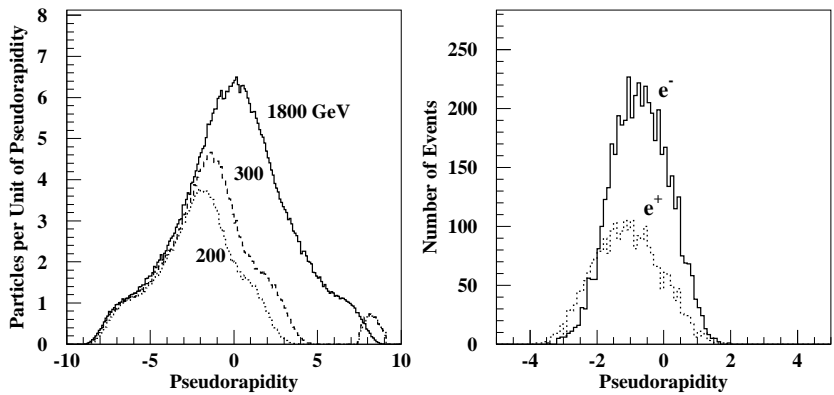,width=5in}}
\vspace*{-2.25in}
\caption{Monte Carlo generated {\em (left)} underlying event and {\em (right)}
lepton-$\eta$  distributions for $p+\bar{p}\rightarrow p+X$; the {\em recoil}
$p$ goes in the positive-$\eta$ direction.}
\label{W}
\end{figure}

A search for diffractive $W's$ is currently being conducted in the CDF
experiment at the Tevatron at $\sqrt{s}=$1800 GeV.  In the absence of a
roman pot spectrometer, the rapidity gap technique is used to tag diffraction.
For non-diffractive (ND) $W(\rightarrow l\nu )$ events, the underlying event
is expected to be right-left symmetric in $\eta -$space (1800 GeV curve
in Fig.~\ref{W}{\em -left}),
while for diffractive
events an asymmetry is expected
($M_X=200$ and 300 GeV curves in Fig.~\ref{W}{\em -left}).
Moreover, the
diffractive asymmetry is correlated with the sign of the lepton-$\eta$ and,
independently, with the sign of the lepton charge.  The
lepton-$\eta$ correlation with the rapidity gap is the result of the
diffractive kinematics, while the charge-gap correlation
is due to the fact that, because of the high $W$ mass, mainly valence quarks
from the quark-flavor asymmetric
(anti)proton interact with the flavor-symmetric pomeron.  The distributions
shown in Fig.~\ref{W}{\em (right)}
were obtained using the POMPYT Monte Carlo program
\cite{POMPYT} with a hard-quark pomeron structure function. Monte
Carlo studies have shown further that non-diffractive
events with a rapidity gap caused by fluctuations in the underlying event
multiplicity do not show any correlations between the gap and the angle or
the charge of the lepton.

\begin{figure}[h,t,b]
\vspace*{-0.5in}
\centerline{\psfig{figure=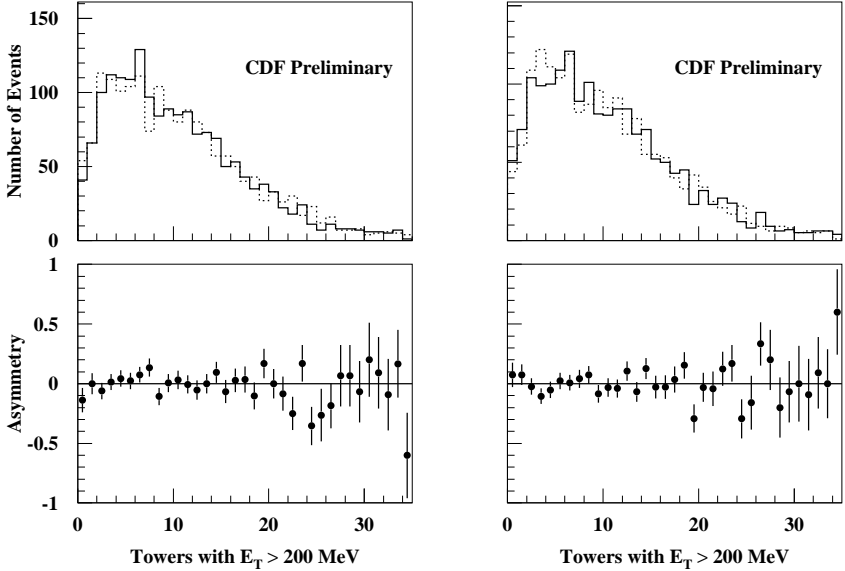,width=5in}}
\vspace*{-1.25in}
\caption{
Charge (left) and angle (right) correlated multiplicity distributions
(solid lines) in the region $2.0<|\eta|<4.2$ compared with the anti-correlated
distributions (dashed lines) for $W$ events;
{\em (bottom)} The asymmetry $A=(N_c-N_{\bar c})/(N_c+N_{\bar c})$
as a function of multiplicity.}
\label{WW}
\end{figure}

Experimentally,
one looks for events devoid of calorimeter towers with transverse energy
$E_T>200$ MeV in the region $2<|\eta|<4.2$, where 4.2 is the maximum
instrumented $\eta$-value in the CDF detector.  Fig.~\ref{WW}{\em (top)}
shows two superimposed
tower-multiplicity distributions in the region
$2<|\eta|<4.2$. The solid histogram is the {\em correlated} multiplicity,
in which the rapidity gap
is opposite the lepton$-\eta$ or for which the (anti)proton
is opposite the $(l^+)l^-$, while the dashed histogram represents the
{\em anticorrelated} multiplicity,
which has opposite to the above correlations.
The events with zero multiplicity (first bin) are diffractive candidates.
The fraction of diffractive events among these candidates
is evaluated
by comparing the two distributions.  Quantitatively, this fraction is
equal to the measured
asymmetry in the first bin,
$A=(N_c-N_{\bar c})/(N_c+N_{\bar c})$,
divided by the asymmetry predicted by the Monte Carlo simulation for
diffractive events (see Fig.~\ref{W}{\em -right}).
This procedure assumes that the asymmetry
expected for non-diffractive events in the first bin is zero.
That this is the case was
verified by Monte Carlo studies, as mentioned above, and can also be inferred
from Fig.~\ref{WW}({\em bottom}), which shows
the asymmetries for all multiplicity bins.

In a sample of $\sim$3,500 $W$ events analyzed, a fraction of $R_{gap}=
(5.8\pm 0.4)$\%
have a rapidity
gap in the region $2<|\eta|<4.2$, but the asymmetry analysis shows that
only a fraction of
$R^D_{gap}=(0.2\pm 0.8)$\% of the events can be attributed to diffractive
production.  To obtain the fraction of all diffractive events in the sample,
$R^D$, the value of
$R^D_{gap}$ must be divided by 0.87 to account for calorimeter noise and by
the relative acceptance of diffractive events
with a gap to all ND events.  While the acceptance, which depends on the
procedure used to model the underlying event,  is currently being evaluated,
it will only affect the uncertainty in the measurement as no signal
has been observed.  Current
indications are that $R^D\sim 0\pm\;a\;few$ \%,
which should be compared with the BI prediction of 17\% and the renormalized
flux prediction of 2.8\%.  An experimental
limit of {\em a few \%} restricts the hard-quark
structure function of the pomeron for the BI-type flux, but
lacks the sensitivity needeed to probe the pomeron structure
if the renormalized flux factor is used.

\subsection{Diffractive dijets in CDF}
The rapidity gap method was also used in CDF to search for diffractive
dijet production, which, as in the UA8 experiment, is sensitive to both
the quark and the gluon content of the pomeron.
Because of the higher energy of the Tevatron,
dijets in the same diffractive mass-region as UA8,
$M_X^2\sim 150$ GeV$^2$, are produced with
lower pomeron $\xi$, since $\xi \approx
M_X^2/s$.  The signature for such events is two high-$P_T$ jets
on the same side of the rapidity region and a rapidity gap on the other
side.  Since the rapidity gap method integrates over $t$, and because of the
exponential $t$-behavior of the diffractive cross section, the average
$t$-value of the events in CDF is close to zero, in contrast to UA8 for
which $|t|\sim 1.5$ GeV$^2$.  Probing the structure of the pomeron with
the same hard process but different pomeron $\xi$ and $t$ can address
the question of the uniqueness of the pomeron structure.
\begin{figure}[h,t,b]
\vspace*{-0.3in}
\centerline{\psfig{figure=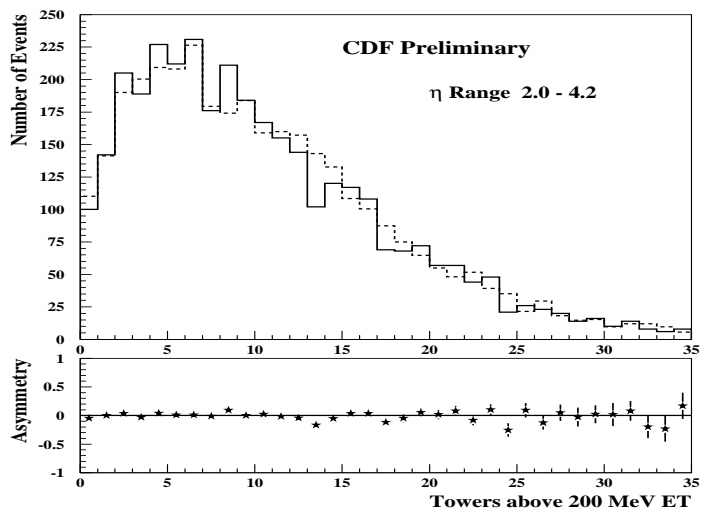,height=3in,width=4in}}
\caption{{\em (top)} Multiplicity distribution in the region $2.0<|\eta|<4.2$
opposite the dijet (solid line) or the lepton from W-decay (dashed line);
{\em (bottom)} The asymmetry $A=(N_{jj}-N_W)/(N_{jj}+N_W)$ as a function of
multiplicity.}
\label{jet}
\end{figure}

A sample of 3415 events with two jets of $P_T>20$ GeV and
$|\eta|>1.8$ was analysed.  As in the $W$ analysis, one looks for
events devoid of calorimeter towers with $E_T>200$ MeV in the region
$2<|\eta|<4.2$.  The multiplicity distribution for the region
{\em opposite} the dijet for all events is plotted in Fig.~\ref{jet}{\em (top)}
and compared with
the same distribution for the W-sample normalized to the same number of
events.  The two distributions agree in all multiplicity bins, including the
bin of zero multiplicity, as shown quantitatively in
Fig.~\ref{jet}{\em (bottom)}.  The
$W$-distribution has two entries per event, one for each side of the rapidity
region. Therefore, the measured fraction of
$(0.2\pm 0.8)$\% diffractive W-candidates in this sample
goes down to $(0.1\pm 0.4)$\% diffractive gaps,
and since there are approximately twice as many W as dijet entries,
the number of diffractive rapidity gap candidates in the
W-sample becomes $(0.05\pm 0.2)$\% of the dijet events.
The dijet-sample contains 3\% rapidity gap events, but the excess of
such events over the W-sample is found to be
$-10\pm 13$, which corresponds to
$(-0.30\pm0.37)$\% of the events.  Because of the negative value obtained,
even if there were diffractive rapidity gap events at the level of the 0.2\%
uncertainty estimated above, correcting for them would  bring the measured
value closer to zero but would not affect the upper limit set by the error
of $\pm 0.37$\%.  The 95\% CL upper limit obtained for the ratio of
diffractive to ND dijets with
a rapidity gap is $R_{gap}<0.6$\%,
which corrected for calorimeter noise and for the relative diffractive to
ND acceptance becomes $R\left(\frac{D}{ND}\right)<0.75$\%.
This value is to be compared with $\sim 5$\%
predicted from  POMPYT for standard flux diffractive and from PYTHIA for
ND dijets, and with
$\sim 0.6$\% obtained using
the renormalized flux.  The corresponding limits on the
discrepancy factor for a full hard-gluon pomeron structure function is
0.14 for the standard flux, and 1.2 for the renormalized flux.  Again, while
this measurement limits  severely the hard-gluon pomeron structure function
for the standard flux, at this level of accuracy it is not a sensitive
probe of the pomeron structure if the renormalized flux is used.

\section{Deep inelastic diffraction at HERA}
At HERA, the quark content of the pomeron is probed directly
with virtual high-$Q^2$ photons in $e^-p$ deep inelastic scattering at
$\sqrt{s}\sim 300$ GeV (28 GeV electrons on 820 GeV protons).
Both the H1 \cite{H1} and ZEUS \cite{ZEUS} Collaborations have
reported measurements of
the diffractive structure function $F^D_2(Q^2,\xi,\beta)$
(integrated over $t$, which is not measured),
where $\beta$
is the fraction of the pomeron's momentum carried by the quark being struck.
The experiments find
that the $\xi$-dependence factorizes out and has the form
$1/\xi^{1+2\epsilon}$, which is the same as the expression in the
pomeron flux factor (see Eq.~\ref{diffractive}).
Moreover, the fits yield $\epsilon\approx 0.1$, which is in agreement
with the value measured in {\em soft}
collisions.

In this paper we evaluate the pomeron structure function
from the H1 results using the renormalized pomeron
flux  \cite{KG3} (a similar analysis could be done on the ZEUS results).
For fixed $Q^2$ and $\beta$, $\xi_{min}=
(Q^2/\beta s)$.  Therefore, the flux integral, which to a good
approximation varies as $\xi_{min}^{-2\epsilon}$, is given by
\begin{equation}
N(\xi_{min})=
N(s,Q^2,\beta)\approx \left(\frac{\beta s}{Q^2}\;\xi_0\right)^{2\epsilon}=
3.8\left(\frac{\beta}{Q^2}\right)^{0.23}
\label{N}
\end{equation}
where $\xi_0$ is the value of $\xi_{min}$ for which the flux integral is unity.
For our numerical evaluations we use $\sqrt{s}$=300 GeV and the flux factor
of Ref.~\cite{KG2}, in which $\epsilon=0.115$.
The value of $\xi_0$ turns out to be $\xi_0=0.004$.

\begin{figure}[h,t,b]
\vspace*{-0.35in}
\centerline{\psfig{figure=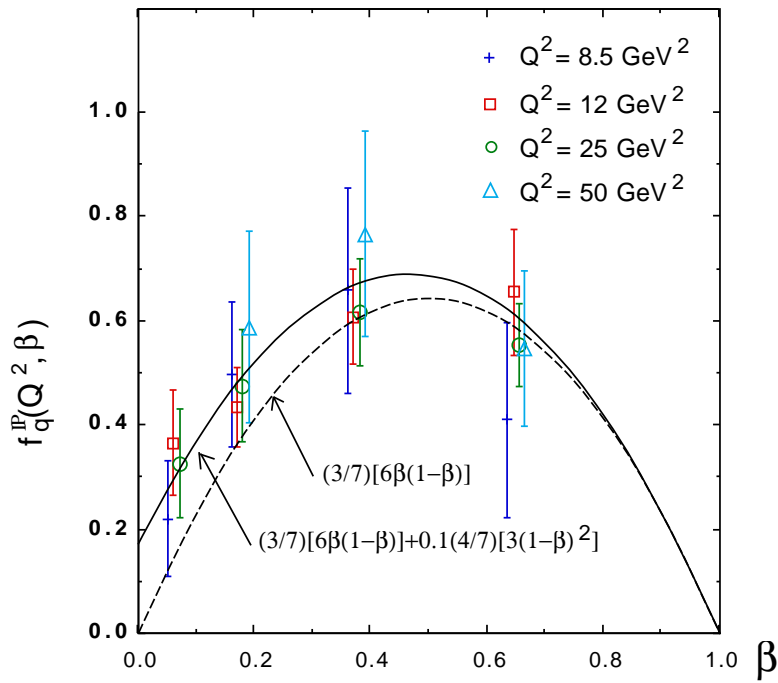,width=5in}}
\vspace*{-2.5in}
\caption{The quark compoment of the pomeron seen in DIS is compared to the
prediction (solid line) based on four quark flavors
and a pomeron that obeys the momentum sum rule; the dashed line represents
the direct quark contribution.}
\label{F2}
\end{figure}

H1 integrates the diffractive form factor
$F^D_2(Q^2,\xi,\beta)$ over $\xi$ and provides values for the expression
\begin{equation}
\tilde{F}^D_2(Q^2,\beta)=\int_{0.0003}^{0.05}F^D_2(Q^2,\xi,\beta)d\xi
\label{F-tilde}
\end{equation}
The pomeron structure function is related to $\tilde{F}^D_2(Q^2,\beta)$ by
factorization:
\begin{equation}
\tilde{F}^D_2(Q^2,\beta)=\left[\frac{\int_{0.0003}^{0.05}d\xi \int_0^{\infty}
f_{{\cal{P}}/p}(\xi,t)\;dt}{N(s,Q^2,\beta)}\right]\,F_2^{{\cal{P}}}(Q^2,\beta)
\label{F2P}
\end{equation}
The expression in the brackets is the normalized flux factor. The integral
in the numerator has the value 2.0 when the flux factor of \cite{KG2} is used.
Assuming now that the pomeron structure function receives contributions
from the four lightest quarks, whose average charge squared is 5/18, the
quark content of the pomeron is given by
\begin{equation}
f^{{\cal{P}}}_{q}(Q^2,\beta)=\frac{18}{5}\,F_2^{{\cal{P}}}(Q^2,\beta)
\label{FQ}
\end{equation}
The values of $f^{{\cal{P}}}_q(Q^2,\beta)$ obtained in this manner are
shown in Fig.~\ref{F2}.

As seen, the renormalized points show no $Q^2$
dependence. We take this fact as an indication that
the pomeron {\em reigns in the kingdom of asymptopia}
and compare the data points with the asymptotic
momentum fractions expected for any quark-gluon construct
by leading-order perturbative QCD, which for
$n_f$ quark flavors are
\begin{equation}
f_q=\frac{3n_f}{16+3n_f}\hspace*{0.5in} f_g=\frac{16}{16+3n_f}
\label{flavors}
\end{equation}
The quark and gluon components of the pomeron structure
are taken to be
 $f_{q,g}^{{\cal{P}}}(\beta)=f_{q,g}\;[6\beta(1-\beta)]$.
For $n_f=4$, $f_q=3/7$ and $f_g=4/7$. The pomeron in this picture is
a combination of valence quark and gluon color singlets and its complete
structure function, which obeys the momentum sum rule, is given by
\begin{equation}
f^{{\cal{P}}}(\beta)=\frac{3}{7}[6\beta(1-\beta)]_q+
\frac{4}{7}[6\beta(1-\beta)]_g
\label{fP}
\end{equation}
The data in Fig.~\ref{F2} are in
reasonably good
agreement with the quark-fraction of the structure function given by
$f_{q}^{{\cal{P}}}(\beta)=(3/7)[6\beta(1-\beta)]$, except for
a small excess at the low-$\beta$ region.  An excess at low-$\beta$
is expected in this
picture to arise from
interactions of the photon
with the gluonic part of the pomeron through gluon splitting into
$q\bar q$ pairs.  Such interactions, which are expected to be down by an order
of $\alpha_s$, result in an {\em effective} quark $\beta$-distribution
of the form $3(1-\beta)^2$. We therefore compare in Fig.~\ref{F2}
the data with the
distribution
\begin{equation}
f_{q,eff}^{{\cal{P}}}(\beta)=(3/7)[6\beta(1-\beta)]+\alpha_s(4/7)[3(1-\beta)^2]
\label{fqeff}
\end{equation}
using $\alpha_s=0.1$. Considering that this distribution involves
{\em no free parameters}, the agreement with the data is remarkable!

\section{Summary and conclusions}
We have reviewed the experimental measurements on hard diffraction
at hadron colliders and on deep inelastic scattering
with large rapidity gaps at HERA.
Using the standard pomeron flux,
the quark component of the pomeron at HERA has a rather flat
$\beta$-distribution \cite{H1,ZEUS} and
integrates out to an average
value of $\bar{f}_q\sim 1/3$. In contrast, UA8 finds a hard structure
with a small amount of soft component, if any;  also, a 1/3 quark component
would almost saturate the UA8 rate, leaving little room for a gluon
component in the pomeron.
Coming now to the CDF results, with such a structure one would
predict a diffractive W rate of $\sim 6-8$\%, depending on the
flux parametrization, which is to be compared with a null result of {\em
a few \%} accuracy.  Thus, the standard flux presents a picture of a mostly
quark-made pomeron with a different momentum sum rule discrepancy factor
for HERA, UA8 and CDF.

Flux renormalization restores order by presenting us with a pomeron that
obeys the momentum sum rule and satisfies all present experimental
constraints.  This pomeron consists of a combination of valence
quark and gluon color
singlets in a ratio suggested by asymptopia for four quark flavors.
In detail, the results obtained with this model are:
\begin{itemize}
\item No free parameters are needed to fit the HERA data (see Fig.~\ref{F2}).
\item HERA and UA8 both find a predominantly
hard structure with a small soft component, which can be accounted for by
gluon-splitting into $q\bar q$ pairs
or gluon radiation by the quarks of the pomeron.
\item For a pomeron consisting of 3/7 quark and 4/7 gluon hard components, the
discrepancy factor for UA8 becomes $1.19\pm 0.18\pm 0.61$, which is
consistent with unity and therefore in agreement with the momentum sum rule.
\item The diffractive $W$ production fraction at the Tevatron
is predicted to be 1.2\%.  This value
is not in conflict with the CDF null result of {\em a few} \% accuracy.
\item The diffractive dijet fraction
at the Tevatron for jet $|\eta|>1.8$ and
$P_T>20$ GeV is predicted to be $\sim 0.5$\%, which is also not in conflict
with the CDF measurement.
\end{itemize}
In conclusion, the pomeron structure function given by Eq.~(\ref{fP})
accounts for
all present experimental results when used in conjunction with the
renormalized flux of \cite{KG2}.

\end{document}